\documentclass[twocolumn,pra,nobalancelastpage,aps,raggedbottom]{revtex4-2}
\usepackage{amsmath,amsfonts,amssymb,graphicx,hyperref,epstopdf,xcolor,tikz,scalerel,natbib}
\usepackage{mathptmx,textcomp}
\usepackage[T1]{fontenc}
\usepackage[utf8]{inputenc}

\newcommand\inlineeqno{\stepcounter{equation}\ (\theequation)}

\definecolor{lime}{HTML}{A6CE39}

\usetikzlibrary{svg.path}
\definecolor{orcidlogocol}{HTML}{A6CE39}
\tikzset{
  orcidlogo/.pic={
    \fill[orcidlogocol] svg{M256,128c0,70.7-57.3,128-128,128C57.3,256,0,198.7,0,128C0,57.3,57.3,0,128,0C198.7,0,256,57.3,256,128z};
    \fill[white] svg{M86.3,186.2H70.9V79.1h15.4v48.4V186.2z}
                 svg{M108.9,79.1h41.6c39.6,0,57,28.3,57,53.6c0,27.5-21.5,53.6-56.8,53.6h-41.8V79.1z M124.3,172.4h24.5c34.9,0,42.9-26.5,42.9-39.7c0-21.5-13.7-39.7-43.7-39.7h-23.7V172.4z}
                 svg{M88.7,56.8c0,5.5-4.5,10.1-10.1,10.1c-5.6,0-10.1-4.6-10.1-10.1c0-5.6,4.5-10.1,10.1-10.1C84.2,46.7,88.7,51.3,88.7,56.8z};
  }
}
\newcommand\orcidicon[1]{\href{https://orcid.org/#1}{\mbox{\scalerel*{
\begin{tikzpicture}[yscale=-1,transform shape] \pic{orcidlogo}; \end{tikzpicture}
}{|}}}}  

\graphicspath{{pics/}}
\DeclareMathAlphabet{\mathpzc}{OT1}{pzc}{m}{it}
\hypersetup{colorlinks,    citecolor=blue,    filecolor=black,    linkcolor=blue,    urlcolor=blue}
\begin{document}
\title{Triple differential cross-section for the twisted electron impact ionization of water molecule}
\author{{Nikita Dhankhar$^1$ \orcidicon{0000-0001-9423-4796}}}


\author{{R. Choubisa$^1$ \orcidicon{0000-0003-3000-6174}} }
\email{rchoubisa@pilani.bits-pilani.ac.in}
\affiliation{$^1$ Department of Physics, Birla Institute of Technology and Science-Pilani, Pilani Campus, Pilani,  Rajasthan, 333031, India}

\begin{abstract}
In this communication, we present the results of the triple differential cross-section (TDCS) for the (e, 2e) process on $H_2O$ molecule for the plane wave and the twisted electron beam impact. The formalism is developed in the first Born approximation. We describe the plane/twisted wave, plane wave, the linear combination of atomic orbitals (LCAO) (self-consistent field LCAO method) and Coulomb wave for the incident electron, scattered electron, the molecular state of $H_2O$, and the ejected electron, respectively. We investigate the angular profiles of the TDCS for the outer orbitals $^1B_1$,$^3A_1$, $^1B_2$ and $^2A_1$ of the water molecule. We compare the angular profiles of the TDCS for the different values of orbital angular momentum (OAM) number {\it m} of the twisted electron beam with that of the plane wave beam. We also study the TDCS for macroscopic $H_2O$ target to explore the effect of opening angle $\theta_p$ of the twisted electron beam on the TDCS. Our results clearly show the effect of the twisted electron’s OAM number ({\it m}) and the opening angle $\theta_p$ on the TDCS of the water molecule.
\end{abstract}

\maketitle

\section{Introduction}
One of the most important collision processes in atomic and molecular physics is the ionization of a given target by an electron impact (hereafter referred as an (e,2e) process). In a coincident (e,2e) process, the interaction of the incident electron leads to the ionization of the target. The ejected and the scattered electrons are detected with their angles and momenta fully resolved \cite{Camp2018}. The study of a coincident (e,2e) process in different kinematical domains is important for investigating the collision dynamics. The (e,2e) study also helps in probing electron correlation and  the structure of a given target \cite{Whelan1992}. The electron impact ionization studies of atoms/molecules have applications in other fields, {\it e.g.} astrophysics, lasers, plasma physics, and radiation physics.  The triple differential cross-section (TDCS) provides detailed information about an (e,2e) process and can be defined as the probability of detecting the outgoing electrons in coincidence with their momenta fully resolved.  The exploration of TDCS for different atomic and molecular targets has progressed significantly with experimental and reliable theoretical results \cite{Whelan1992,Colgan2002,LB2009,Al2010,XRen2015,Mouawad2017,Khatir2019,GPurohit2021,Lozano2021}.

	Water molecule plays a crucial role in biological matters. The study of ionization of water molecule is valuable in radiology, radiation treatment and planetary atmosphere \cite{2000Bouda,2015Blanc,2015Ali}. To dig into the finer aspects of charged particle interactions in a biological medium, the study of ionizing processes by electron impact for water molecules is important. 
The calculation of differential cross-sections for the  molecular (e,2e) process is challenging when compared to that for the atomic (e,2e) process because of the complex molecular configurations and the orientation dependency of the molecule.	
In literature, various theoretical models have been employed for the study of (e,2e) processes on water molecule. Champion {\it et al.} \cite{Champion2006} used different models, such as the one Coulomb wave (1CW) (using the partial wave expansion method), the distorted wave Born approximation (DWBA), the Brauner Briggs Klar (BBK), the two Coulomb wave (2CW) and the dynamic screening of the three two-body Coulomb interactions (DS3C), to study the (e,2e) process on $H_2O$ molecule. 1CW model with Gaussian type orbitals (GTO) has also been used to study the TDCS \cite{Champion2009}. Different other models, like 1CW (analytical expression) \cite{Sahlaoui2011, Sanctis2015}, generalised Sturmian function (GSF) \cite{Castro2016}, two molecular three-body distorted wave approach (M3DW) \cite{Ren2017}, multicenter three distorted waves (MCTDW) \cite{Gong2018}, second-order distorted wave Born approximation (DWBA2)\cite{Psingh2019}, have also been used.
In addition to (e,2e) processes, various theoretical (e,3e) investigations have been done  to study the double ionization on $H_2O$ \cite{Champion2010,Jones2010,Oubaziz2015}.


So far, only plane wave electron beams (with no orbital angular momentum (OAM)) have been used to study the (e,2e) collision experiments on $H_2O$.
The experimental realization of electron vortex beams (also known as twisted electron beams, carrying an additional OAM) by different groups ushered in a new era of research into investigating the interactions of atomic and molecular targets with twisted electron beams \cite{Uchida2010, Verbeeck2010, Morran2011}. 
The term ``vortex beam'' here refers to a freely propagating electron beam having a helical wave-front and a well-defined OAM, {\it m}, along the propagation direction. These beams have a helical phase front $e^{i m\phi}$ with the azimuthal angle $\phi$ about the propagation axis (for more details about EVBs, see \cite{Llyod2017, Bliokh2017, Hugo2018}).
Twisted electron beams provide scope for research in optical microscopy, quantum state manipulation, optical tweezers, astronomy, strong-field ionization, and many more fields \cite{Verbeeck2010, Roe2015, Qiu2018, Max2021}.

	The intrinsic angular momentum of the electron vortex beam influences the role of an electron in the ionization process \cite{Boxem2016}. Therefore, it is essential to understand the interaction of electron beam, with non-zero OAM, at atomic/molecular level to explore their applications to other fields. Theoretical descriptions of radiative recombination, elastic scattering, impact ionization, and impact excitation have been investigated so far. 
	The work by Ivanov and Serbo (2011) \cite{Serbo2011}, Boxem {\it et al.} (2014) \cite{Boxem2014} and   Boxem \textit{et al.} \cite{Boxem2015} contributed to the outset of the theoretical analysis of the scattering experiments by twisted electrons. Serbo \textit{et al.} \cite{Serbo2015} analyzed the scattering by twisted electrons in the relativistic framework.  Sch\"uler and Berakdar theoretically investigated the Electron Energy-Loss Spectroscopy (EELS) for the $C_{60}$ fullerene target \cite{Schuler2016} by a twisted electron beam. Karlovets \textit{et al.} \cite{Karlovets2017} studied the scatterig in the framework of Born approximation. 
Maiorova \textit{et al.} \cite{Maiorova2018} advanced the scattering studies by their theoretical analysis of the Differential Cross-Section (DCS) for the elastic scattering of twisted electrons by molecular hydrogen $H_2$. The angular distribution of the DCS highlighted the influence of the twisted electron beam's parameter on the Young-type interference. Harris \textit{et al.} (2019) reported the ionization of the Hydrogen atom by twisted electrons by analyzing the fully differential cross-section (FDCS) with different parameters of the twisted electron beam \cite{Harris2019}. The results indicated a shift in the binary and recoil peak for twisted electron cross-sections from their plane wave locations due to the projectile’s transverse momentum components. The recent study by Mandal \textit{et al.} (2020) showed the dependence of the Total Angular Momentum (TAM) number (\textit{m}), and opening angle ($\theta_p$) on the angular profile of the TDCS and spin asymmetry for the relativistic electron impact ionization of the heavy atomic targets \cite{Mandal2020}. Furthermore, Dhankhar \textit{et al.} \cite{Dhankhar2020}  studied theoretically the double ionization of He atom in $\theta$-variable and constant $\theta_{12}$ mode for the twisted electron incidence. The same group also investigated the Five-Fold Differential Cross-section (FDCS) and TDCS for the single ionization of molecular hydrogen ($H_2$). They also explored the influence of the twisted electron beam on the (e,2e) process on the $H_2$ molecule from the perspective of the ‘Young-type’ interference of the scattered waves emanating from the two atomic centers of the $H_2$ molecule. Their results found that the angular profile of the TDCS and FDCS depends on the OAM number, \textit{m}, and the opening angle, $\theta_p$, of the incident twisted electron beam \cite{Dhankhar2020_2}.
 
To our knowledge, the investigations on electron impact ionization by twisted electron beams to date have been performed for single ionization of the hydrogen-like target atoms and $H_2$ molecule. In this communication, we present the first theoretical estimation for the ionization of water molecule for the twisted electron beam. Our study is performed within the first Born approximation framework for an incident plane wave electron beam (Sec.\ref{sec2a}) and for an incident twisted electron beam in Sec.\ref{sec2b}. We describe the plane wave, Slater-type wave-functions, Coulomb wave for the scattered electron, the molecular state of $H_2O$, and the ejected electron respectively. We present our results of the TDCS for the outer orbitals, namely, $^1B_1$, $^1B_2$, $^3A_1$ and $^2A_1$, of the water molecule for different parameters of the twisted electron beam in Sec.\ref{sec3}. Finally, we conclude our paper in the Sec.\ref{sec4}. Atomic units are used throughout the paper unless otherwise stated.

\section{Theoretical Formalism}\label{sec2}
This section presents the theoretical formalism for the computation of (e,2e) differential cross-sections of a water molecule for both the plane wave and the twisted electron beam as an incident beam.
 
	The (e,2e) process on a water molecule can be described as;
\begin{equation}\label{1}
e_i^- + H_2O \rightarrow e_s^- + e_e^- + H_2O^+
\end{equation}
here, $e_i^-$, $e_s^-$ and $e_e^-$ represents the incident, scattered and ejected electron respectively.

In the present theory, we apply the closure relation over all the possible rotational and vibrational states of the residual target ($H_2O^+$ ion). Thus the electron impact ionization of water molecule considered here is a pure electronic transition. We have neglected the exchange effects between the incident/scattered and bound/ejected electron since the incident/scattered electron is faster than the bound/ejected electron for the energies considered here\cite{Champion2006}.

\subsection{Plane wave ionization cross-sections}\label{sec2a}

\begin{figure}
\includegraphics[width = 0.85\columnwidth]{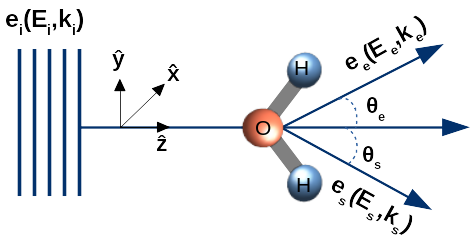}
\caption{Schematic diagram for the electron impact ionization of $H_2O$ molecule, centered at oxygen nucleus, by plane wave in co-planar asymmetric geometry. An incident plane wave of energy $E_i$ and momentum $\mathbf{k}_i$ interacts with the $H_2O$ molecule ejecting one of the bound electrons of the molecule into continuum state. We describe $E_s$ and $E_e$ and $\mathbf{k}_s$ and $\mathbf{k}_e$ as the energies and momenta of the scattered and ejected electron respectively. $\theta_s$ and $\theta_e$ represent the angular positions of the scattered and ejected electrons. The {\it z}-axis is chosen along the propagation direction of the plane wave. The scattering plane is the {\it xz}-plane and the scattered and ejected electrons are detected in scattering plane ($E_s > E_e$).}\label{fig1}
\end{figure}

	For an (e,2e) process on $H_2O$ molecule, the five-fold differential cross-section (5DCS) of a molecular orbital is given by \cite{Sanctis2015},
\begin{equation}\label{2}
  \begin{aligned}
\sigma^{(5)}(\alpha,\beta,\gamma) & = \frac{d^5\sigma}{d\omega d\Omega_e d\Omega_s dE_e}\\
& =  (2\pi)^4 \frac{k_ek_s}{k_i} |T_{fi}^{pw}|^2,
  \end{aligned}
\end{equation}
where $d\omega = \sin\beta d\alpha d\beta d\gamma$
is the solid angle element for the molecular orientation in the laboratory frame and $\alpha,\beta$ and $\gamma$ are the Euler angles of the water molecule. $dE_e$ describes the energy interval for the ejected electron and $d\Omega_s$ and $d\Omega_e$ are the solid angle's intervals of the scattered and the ejected electron respectively. $\mathbf{k_i}$, $\mathbf{k_s}$ and $\mathbf{k_e}$ represents the momentum of the incident, scattered and ejected electron respectively.
The scattering amplitude, $T_{fi}^{pw}$, describing the transition from the initial state $\psi_i$ to final state $\psi_f$ for a plane wave (in the first Born approximation) is given by,
\begin{equation}\label{4}
T_{fi}^{pw} = \langle \psi_f |V| \psi_i \rangle,
\end{equation}
where V describes the interaction between the incident electron and the molecular $H_2O$ target and is given by;
\begin{equation}\label{5}
V = \frac{-8}{r_0} - \frac{1}{|\mathbf{r}_0-\mathbf{R}_{OH_1}|} - \frac{1}{|\mathbf{r}_0-\mathbf{R}_{OH_2}|} + \sum_{i = 1}^{10} \frac{1}{|\mathbf{r}_0-\mathbf{r}_i|},
\end{equation}
where $\mathbf{R}_{OH_1}$ and $\mathbf{R}_{OH_2}$ represent the position vector of the two hydrogen nuclei from the oxygen nucleus with $|R_{OH_1}|$ = $|R_{OH_2}|$ = 1.814 a.u.. $\mathbf{r}_i$ is the position vector of the $i^{th}$ bound electron of the target with respect to the center of the oxygen nucleus (assumed to be fixed) and $\mathbf{r_0}$ is the position vector of the incident particle\cite{Champion2010}.

	 We develop the theoretical formalism with the following assumptions;
 \begin{enumerate}
\item Both the incident and the scattered electrons are described as a plane wave.
\item The molecular wave-function, $\Phi_j(\mathbf{r})$, is expressed as the linear combinations of the Slater-type functions centered at oxygen nucleus (self-consistent field LCAO \cite{Moccia1964}).
\item The ejected electron is described by a Coulomb wave function, $\psi_{\mathbf{k}_e}^-(\mathbf{r})$.
\item The exchange effects between the incident/scattered electron with the bound/ejected electron are neglected here since the incident/scattered electron is faster than the bound/ejected electron. 
\end{enumerate}
The electronic structure of water molecule consists of ten bound electrons which are distributed among five one-center molecular orbitals expressed by Linear Combination of Atomic Orbitals (LCAO). The orbitals are $^1B_1$, $^3A_1$, $^1B_2$, $^2A_1$ and $^1A_1$. The LCAO of each molecular orbital is characterised by a dominant atomic orbital component . The orbital  $^1B_1$ has 2$p_{+1}$, $^3A_1$ has 2$p_0$, $^1B_2$ has 2$p_{-1}$ , $^2A_1$ has 2$s$ and $^1A_1$ has 1$s$ dominant atomic orbital component \cite{Champion2001}.
The molecular orbitals expressed by the linear combinations of the Slater-type functions are given as (we follow the same mathematical representation as  in \cite{Champion2006});
\begin{equation}\label{6}
\Phi_j(\mathbf{r}) = \sum_{k = 1} ^{N_j}a_{jk} \phi_{n_{jk}l_{jk}m_{jk}}^{\xi_{jk}}(\mathbf{r}),
\end{equation}
where $N_j$ is the number of Slater functions used to describe the $j^{th}$ molecular orbital and $n_{jk}$,$l_{jk}$,$m_{jk}$ are the quantum numbers for the $j^{th}$ molecular orbital. $a_{jk}$ is the weight of each atomic component $\phi_{n_{jk}l_{jk}m_{jk}}^{\xi_{jk}}(\mathbf{r})$ and  $\xi_{jk}$ is a variational paramater.  $\phi_{n_{jk}l_{jk}m_{jk}}^{\xi_{jk}}(\mathbf{r})$ is  expressed as \cite{Champion2005};

\begin{equation}\label{7}
\phi_{n_{jk}l_{jk}m_{jk}}^{\xi_{jk}}(\mathbf{r}) = R^{\xi_{jk}}_{n_{jk}}(r) S_{l_{jk}m_{jk}}(\mathbf{\hat{r}}),
\end{equation}
 where 
 $R^{\xi_{jk}}_{n_{jk}}(r)$ is the radial part of each atomic orbital and given as;
 \begin{equation}\label{8}
R^{\xi_{jk}}_{n_{jk}}(r) = \frac{(2\xi_{jk})^{n_{jk}+\frac{1}{2}}}{\sqrt{2n_{jk}!}}r^{n_{jk}-1}e^{-\xi_{jk}r},
\end{equation}\\
and $S_{l_{jk}m_{jk}}(\mathbf{\hat{r}})$ is the real spherical harmonics expressed as\cite{Helgaker2014},

with $m_{jk} \neq 0$: 
\begin{equation}\label{9}
\begin{aligned}
S_{l_{jk}m_{jk}}(\mathbf{\hat{r}}) ={} & \sqrt{\Big( \frac{m_{jk}}{2|m_{jk}|} \Big)} \Bigg\{  Y_{l_{jk}-|m_{jk}|}(\mathbf{\hat{r}})+ \\
 &  (-1)^m_{jk} \Big( \frac{m_{jk}}{|m_{jk}|} \Big)Y_{l_{jk}|m_{jk}|}(\mathbf{\hat{r}})  \Bigg\},
\end{aligned}
\end{equation}

and $m_{jk} = 0$: \hskip2ex
$S_{l_{jk}0}(\mathbf{\hat{r}}) = Y_{l_{jk}0}(\mathbf{\hat{r}}). \hskip 12ex \inlineeqno$ \label{10}

Here $Y_{lm}(\hat{\mathbf{r}})$ is the complex spherical harmonics.
	The linear combination of spherical harmonics can be used for the transformation of the molecular orientation from the molecular frame to the laboratory frame expressed as \cite{Sahlaoui2011}:
\begin{equation}\label{13}
S_{lm}(\mathbf{\hat{r}}) = \sum_{\mu = -l}^{l} D_{m\mu}^{(l)}(\alpha, \beta, \gamma) S_{l\mu}(\mathbf{\hat{r}}),
\end{equation}	
where $D_{m\mu}^{(l)}(\alpha, \beta, \gamma)$ is the rotation matrix with Euler angles $\alpha$, $\beta$ and $\gamma$.

 The problem of N(=10) electrons can be reduced to one active electron problem using the frozen-core approximation. Within the framework of an independent electron approximation, it is assumed that one of the target electrons (the {\it active} one) is ejected in the final channel of the reaction, whereas the other electrons (the passive electrons) remain as frozen in their initial sates \cite{Sanctis2015, Sahlaoui2012}. Performing the integration over $\mathbf{r}_0$ analytically (see \cite{Tweed1992}), we have,
\begin{equation}\label{11}
\int \frac{e^{i\mathbf{K} \cdot \mathbf{r}_0}}{|\mathbf{r}_0 - \mathbf{r}_1|} d^3r_0 = \frac{4\pi}{K^2}e^{i\mathbf{K} \cdot \mathbf{r}_1}. 
\end{equation}
The matrix element $T_{fi}$ is thus given by,
\begin{equation}\label{12}
T_{fi}^{pw}(\mathbf{q}) = \frac{-2}{q^2} \langle \psi^-_{\mathbf{k}_e} | e^{i\mathbf{q}\cdot\mathbf{r}} - 1| \Phi_j(\mathbf{r}) \rangle,
\end{equation}
where $\mathbf{q} = \mathbf{k}_i - \mathbf{k}_s$ is the momentum transferred to the target. Here, $\psi_{\mathbf{k}_e}^-(\mathbf{r})$ and $\Phi_j(\mathbf{r})$ represent the Coulomb wave-function and the molecular wave-function respectively.

For the gas-phase ionization of the water molecule, 
experimentally it is not possible to align the molecule in one particular orientation. Thus we compute the Triple differential cross section (TDCS) by taking an average over all the possible orientations of the water molecule. We obtain the TDCS by integrating 5DCS over the Euler's angle and is given by;

\begin{equation}\label{3}
  \begin{aligned}
\sigma^{(3)}  & = \frac{d^3\sigma}{d\Omega_e d\Omega_s dE_e}\\
& = \frac{1}{8 \pi^2} \int \sigma^{(5)}(\alpha,\beta,\gamma)  \sin \beta d\alpha d\beta d\gamma.
  \end{aligned}
\end{equation}

The integration over the Euler angles can then be performed using the ortho-normalization property of the rotation matrix and the TDCS is then given by \cite{Sahlaoui2011},
\begin{equation}\label{14}
\frac{d^3\sigma}{d\Omega_e d\Omega_s dE_e} = \frac{k_e k_s}{k_i} \sum_{k = 1} ^{N_j} \frac{a_{jk}^2}{\hat{l}_{jk}}\sum_{\mu = -l_{jk}}^{l_{jk}} |T_{fi}^{pw}(\mathbf{q})|^2,
\end{equation}
where $\hat{l}_{jk}$ = $2l_{jk}+1$.

 \begin{figure}[ht]
\includegraphics[width = 1.0\columnwidth]{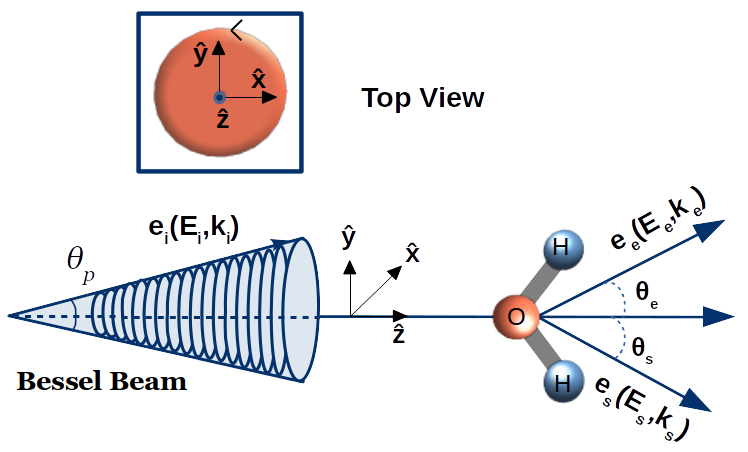}
\caption{Same as figure \ref{fig1}, except the incident beam is the twisted electron beam with an opening angle $\theta_p$ and OAM number ``m''. The quantization (\textit{z})-axis is chosen along the propagation direction of the incoming beam. The inset represents the top view of the incident twisted electron beam. The beam propagates out of the page and twists around the propagation direction (counter-clockwise).} \label{fig2}
\end{figure}

\subsection{Twisted electron ionization cross-sections}\label{sec2b}

\begin{figure*}
\includegraphics[width=18cm,height=12cm]{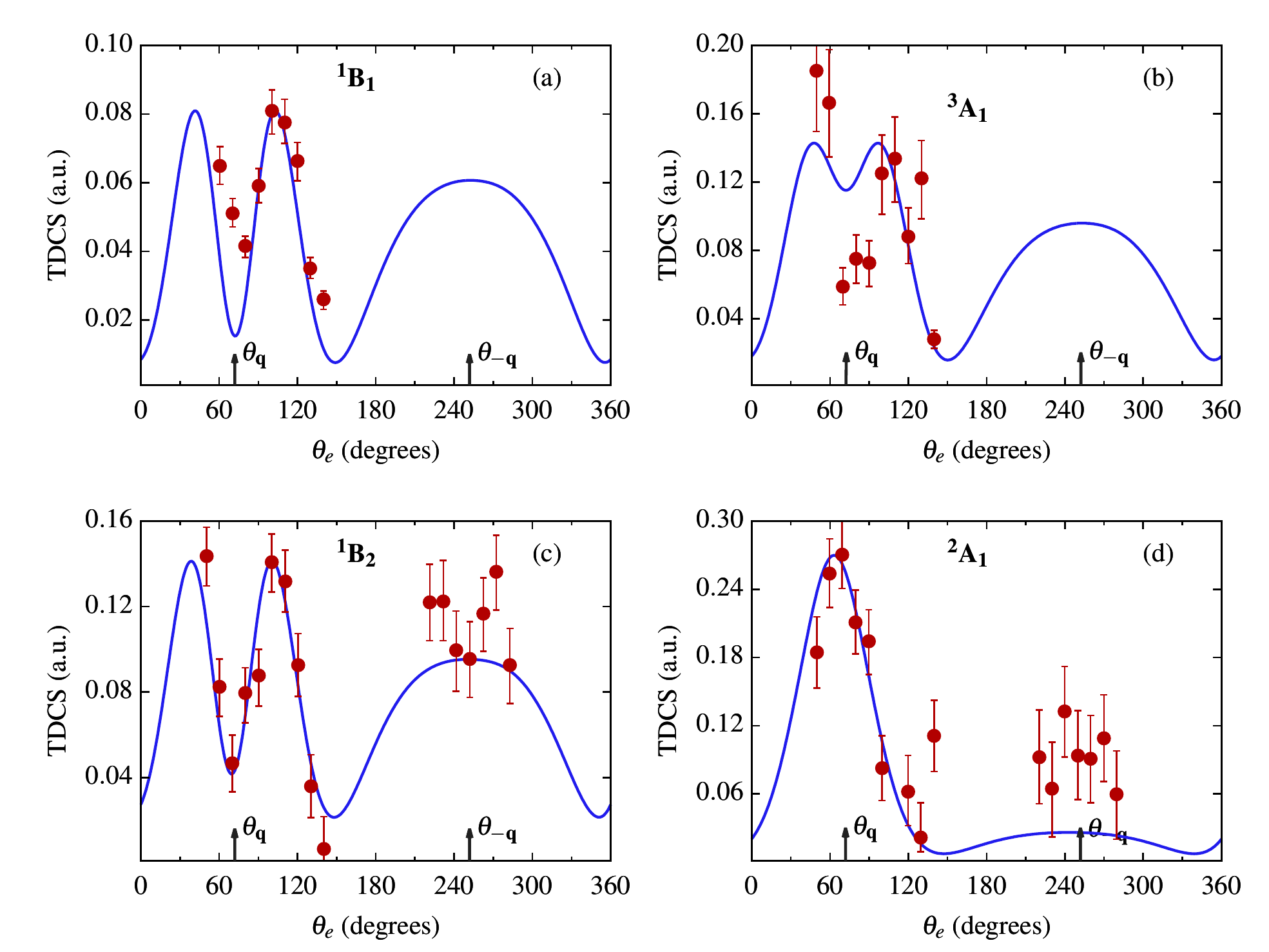}
\caption{TDCS as a function of ejected electron angle $\theta_e$ for the plane wave (e,2e) process on $H_2O$ molecule in the co-planar asymmetric geometry. Our plane wave results are represented by solid line and experimental results (\cite{Milne2004}) by full circles. The kinematics used here is: $E_i$ = 250eV, $E_e$ = 10eV (except 8eV for $^3A_1$ orbital (sub-figure (b))) and $\theta_s$ = 15\textdegree. Arrows indicate the direction of momentum transfer $(\theta_\mathbf{q})$ and recoil direction $(\theta_{-\mathbf{q}})$ (opposite to $(\theta_\mathbf{q})$ direction) for this and all subsequent figures.}
\label{fig3}
\end{figure*}

Figure \ref{fig2} illustrates the (e,2e) process on $H_2O$ molecule by the twisted electron beam. 
A twisted electron beam is characterized by a helical wave-front that twists around the beam axis as the beam propagates along the propagation direction.
We use the same formalism as mentioned in Sec.\ref{sec2a}, for the computation of the TDCS, except here we replace the plane wave function for the incident electron  with a twisted electron wave-function, such as a Bessel beam \cite{Boxem2015}.

	For an incident twisted electron beam, the momentum vector $\mathbf{k}_i$, can be described as \cite{Dhankhar2020}, 
\begin{equation}\label{15}
\mathbf{k}_i = (k_i \sin\theta_p \cos\phi_p)\hat{x} + (k_i \sin\theta_p \sin\phi_p)\hat{y} + (k_i \cos\theta_p)\hat{z}
\end{equation}
where $\theta_p$  and $\phi_p$ are the polar and azimuthal angles of the $\mathbf{k}_i$ respectively. 
The longitudinal momentum is along the \textit{z}-axis. The momentum vector, $\mathbf{k}_i$, forms the surface of a cone with an angle $\theta_p$ with the \textit{z}-axis, which is normally referred as the opening angle of the twisted beam. $\theta_p = \tan^{-1}\frac{k_{i\perp}}{k_{iz}}$ with  $k_{i\perp}$ and $k_{iz}$ are the perpendicular and the longitudinal components of the momentum  $\mathbf{k}_i$ respectively.

	Experimentally it is difficult to obtain an exact alignment of the incident Bessel beam with the target, therefore, one needs to consider the Bessel beam with non-zero impact parameter ($\mathbf{b}$). This gives a more generalized equation for the Bessel beam \cite{Serbo2015} (which can be then be used to compute the TDCS for a macroscopic target by taking an average over all the possible impact parameters), such that
\begin{equation}\label{16}
\psi^{(tw)}_{\varkappa m}(\mathbf{r}_0) = \int^{\infty}_{0} \frac{dk_{i\perp}}{2\pi}\ k_{i\perp} \int_{0}^{2\pi}\frac{d\phi_p}{2\pi}\ a_{\varkappa m}(k_{i\perp})e^{i\mathbf{k}_i \cdot \mathbf{r}_0}e^{-i\mathbf{k}_i \cdot \mathbf{b}},
\end{equation}
where $\varkappa$ is the absolute value of the transverse momentum ($k_i sin(\theta_p)$). $\mathbf{b}$ is the vector that describes the transverse orientation of the incident twisted electron beam with respect to the target. The impact parameter vector $\mathbf{b}$ is described as $\mathbf{b} = b \cos\phi_b \hat{x} + b\sin\phi_b \hat{y}$, with \textit{b} as the magnitude of $\mathbf{b}$ and $\phi_b$ as the azimuthal angle of $\mathbf{b}$. In contrast to plane wave, the additional factor $e^{-i\mathbf{k}_i \cdot \mathbf{b}}$ in equation (\ref{16}), implies the complex spatial structure of the Bessel beam \cite{Serbo2015}.

	By substituting the plane wave-function with the Bessel wave-function (equation(\ref{16}) in equation(\ref{4})), we obtain the twisted wave transition amplitude ($T_{fi}^{tw}(\varkappa,\mathbf{q})$) in terms of the plane wave transition amplitude $T_{fi}(\mathbf{q})$ (equation (\ref{12})) as (in the frozen-core approximation)\cite{Dhankhar2020_2};
\begin{equation}\label{17}
T^{tw}_{fi}(\varkappa,\mathbf{q},\mathbf{b}) = (-i)^m \int_{0}^{2\pi} \frac{d\phi_p}{2\pi}\ e^{im\phi_p - i\mathbf{k}_{i\perp}\cdot\mathbf{b}}\ T_{fi}^{pw}(\mathbf{q}),
\end{equation}
where, $\mathbf{k}_{i\perp}\cdot\mathbf{b}$ = $\varkappa b \cos(\phi_p-\phi_b)$. 
The momentum transfer to the target for a twisted electron beam case can be expressed as; 
\begin{equation}\label{19}
q^2 = k_i^2 + k_s^2 - 2k_ik_s \cos\theta ,
\end{equation}
where, 
\begin{equation}\label{20}
\cos\theta = \cos\theta_p \cos\theta_s + \sin\theta_p \sin\theta_s \cos(\phi_p - \phi_s).
\end{equation}
In the above equation $\theta_s$ and $\phi_s$ are the polar and azimuthal angles of the $\mathbf{k}_s$. For the co-planar geometry $\phi_s = 0$.  

	Here, the molecular target is assumed to be located along the direction of the incident twisted electron beam ({\it z-}axis). Thus by using $\mathbf{b} = 0$ in equation (\ref{17}), the twisted wave transition amplitude $T^{tw}_{fi}(\varkappa,\mathbf{q})$, can be written as, 
\begin{equation}\label{18}
T^{tw}_{fi}(\varkappa,\mathbf{q}) = (-i)^m \int \frac{d\phi_p}{2\pi}\ e^{im\phi_p}\ T_{fi}^{pw}(\mathbf{q}).
\end{equation}
The TDCS for the molecular orbital of water molecule  by twisted electron can be computed from equation (\ref{18}) together with the transition amplitude $T_{fi}^{pw}( \mathbf{q})$ from the equation (\ref{12}).

	The process of ionization of a single molecule by a vortex beam is challenging experimentally. Therefore, in a more realistic scenario the ionization process on a macroscopic target is  preferable \cite{Zaytsev2020}. The cross-section for such a target can then be computed by taking the average of the plane wave cross-sections over all the possible impact parameters, \textbf{b}, in the transverse plane of the twisted electron beam.
The average cross-section, (TDCS)$_{av} = \overline{\frac{d^3 \sigma}{d\Omega_s d\Omega_e dE_e}}$ in terms of plane wave cross-section can be described as (for detailed derivation see  \cite{Serbo2015,Karlovets2017,Harris2019});
\begin{equation}\label{21}
(TDCS)_{av} = \frac{1}{2 \pi \cos\theta_p} \int_{0}^{2\pi} d\phi_p \frac{d^3 \sigma(\mathbf{q})}{d\Omega_s d\Omega_e dE_e},
\end{equation}
where $\frac{d^3 \sigma(\mathbf{q})}{d\Omega_s d\Omega_e dE_e}$ is like the TDCS for the plane wave electron beam depending on \textbf{q}.
From equation (\ref{21}), it is evident that the cross-section for the scattering of the twisted electrons by the macroscopic target is independent of the OAM number {\it m} of the incident twisted electron beam. However, (TDCS)$_{av}$ depends on the opening angle $\theta_p$ of the incident twisted electron beam.

\begin{figure}[!ht]
\includegraphics[width=1.0\columnwidth]{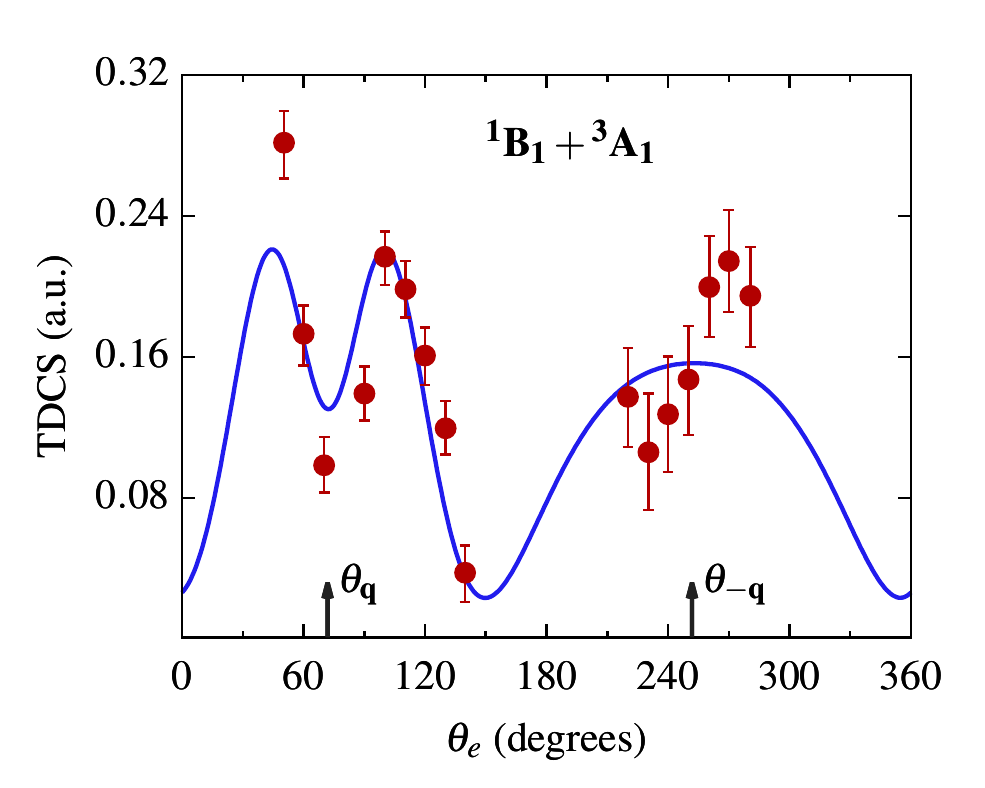}
\caption{Same as figure \ref{fig3}, except the summed TDCSs of the $^1B_1$ and $^3A_1$ orbitals are plotted. Kinematics is same as mentioned in figure \ref{fig3}}
\label{fig4}
\end{figure} 

\section{Results and discussions}
\label{sec3}

In this section, we present the results of our calculations of the TDCS for $H_2O$ by a twisted electron beam. We benchmark our theoretical results with the existing experimental data for the plane wave. We study the effect of different parameters of the twisted electron beam on the angular profiles of (e,2e) cross-sections for different orbitals of the water molecule.  We present the single ionization differential cross-section for the water molecule averaged over orientation in the gaseous phase. We compare our twisted electron beam results with that of plane wave results for different values of orbital angular momentum (OAM) number {\it m}, {\it viz.} 1, 2, and 3. The kinematics we have used here is; incident energy ($E_i$) = 250eV, ejected energy ($E_e$) = 10eV (except for the $^3A_1$ molecular orbital for which $E_e$ = 8eV), $\theta_s$ = 15\textdegree \  in the coplanar asymmteric geometry, similar to Milne-Brownlie {\it et al.}\cite{Milne2004} for the plane wave (e,2e) process. Since, experimentally, it is difficult to align the molecule in a particular direction, we compute the TDCS here.

\begin{figure*}
\includegraphics[width= 2.0\columnwidth]{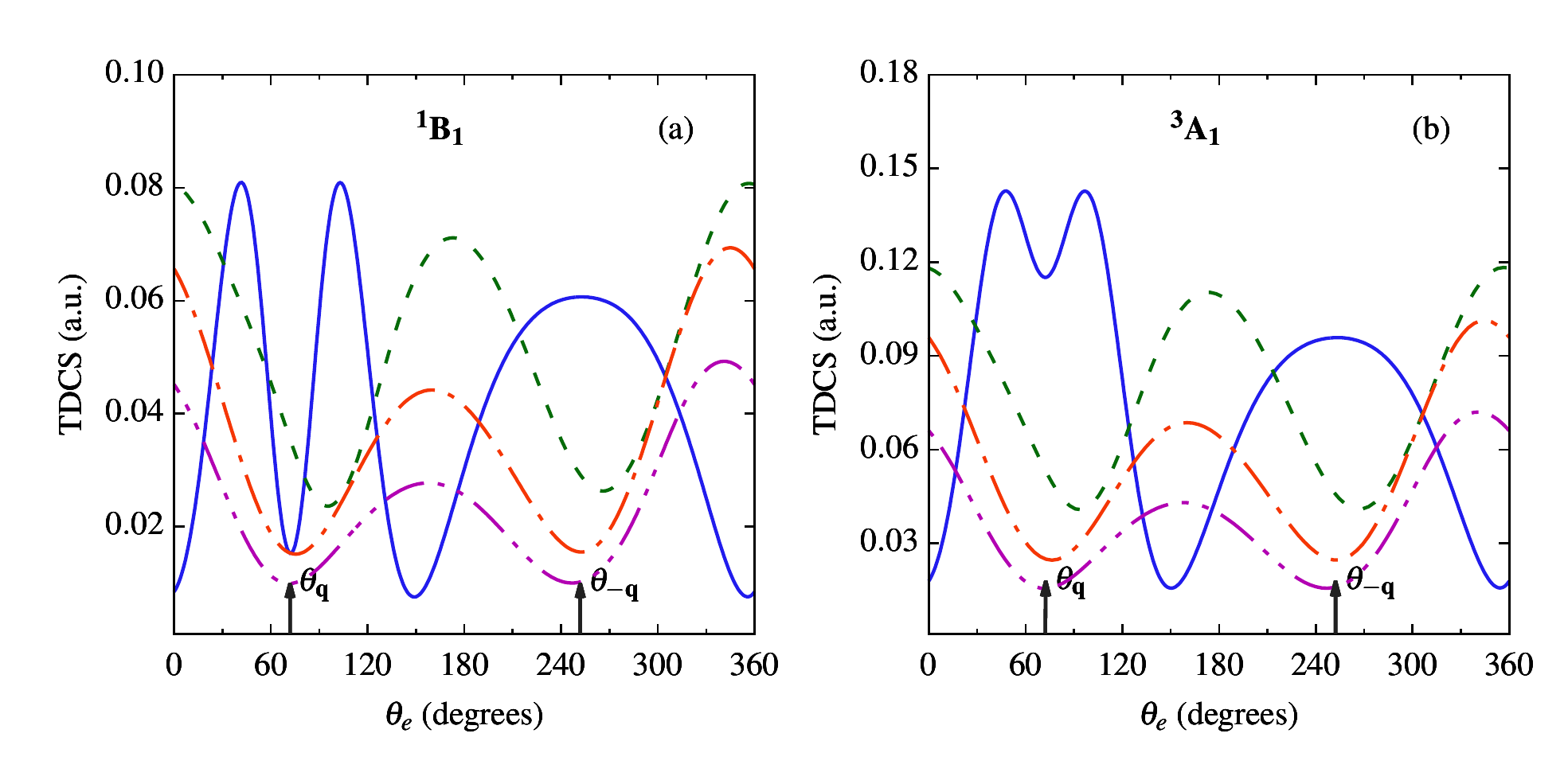}
\caption{TDCS as a function of ejection angle $\theta_e$ for the twisted electron wave (e,2e) process on $H_2O$ molecule in the co-planar asymmetric geometry. The kinematical conditions are $E_i$ = 250eV, $E_e$ = 10eV (except 8eV for $^3A_1$ orbital (sub-figure (b))) and $\theta_p$ = $\theta_s$ = 15\textdegree. The solid, dashed, dashed-dotted and dashed-dotted-dotted curves represent the plane wave, {\it m} = 1, 2 and 3 respectively. The results for {\it m} $\neq$ 0 are scaled up by a factor of 2 for both the sub-figures (a) and (b)}\label{fig5}
\end{figure*}

\subsection{Angular profiles of the TDCS for plane wave}

We present in figure \ref{fig3} the results of our calculations of the TDCS as a function of the ejected electron's angle ($\theta_e$) for the plane wave electron beam in the co-planar asymmetric geometry for the outer orbitals of the water molecule, namely, $^1B_1$, $^3A_1$, $^1B_2$ and $^2A_1$. The arrows in the figure \ref{fig3} and subsequent figures represent the direction of momentum transfer ($\theta_{\mathbf{q}}$) and the recoil direction ($\theta_{-\mathbf{q}}$).  We compare the results of our calculation of the TDCS with the experimental data reported by Milne-Brownlie {et al.} \cite{Milne2004} to benchmark our calculations so that we can validate our theoretical calculations for the twisted electron beam.
 
\begin{figure*}
\centering
\includegraphics[width=2.0\columnwidth]{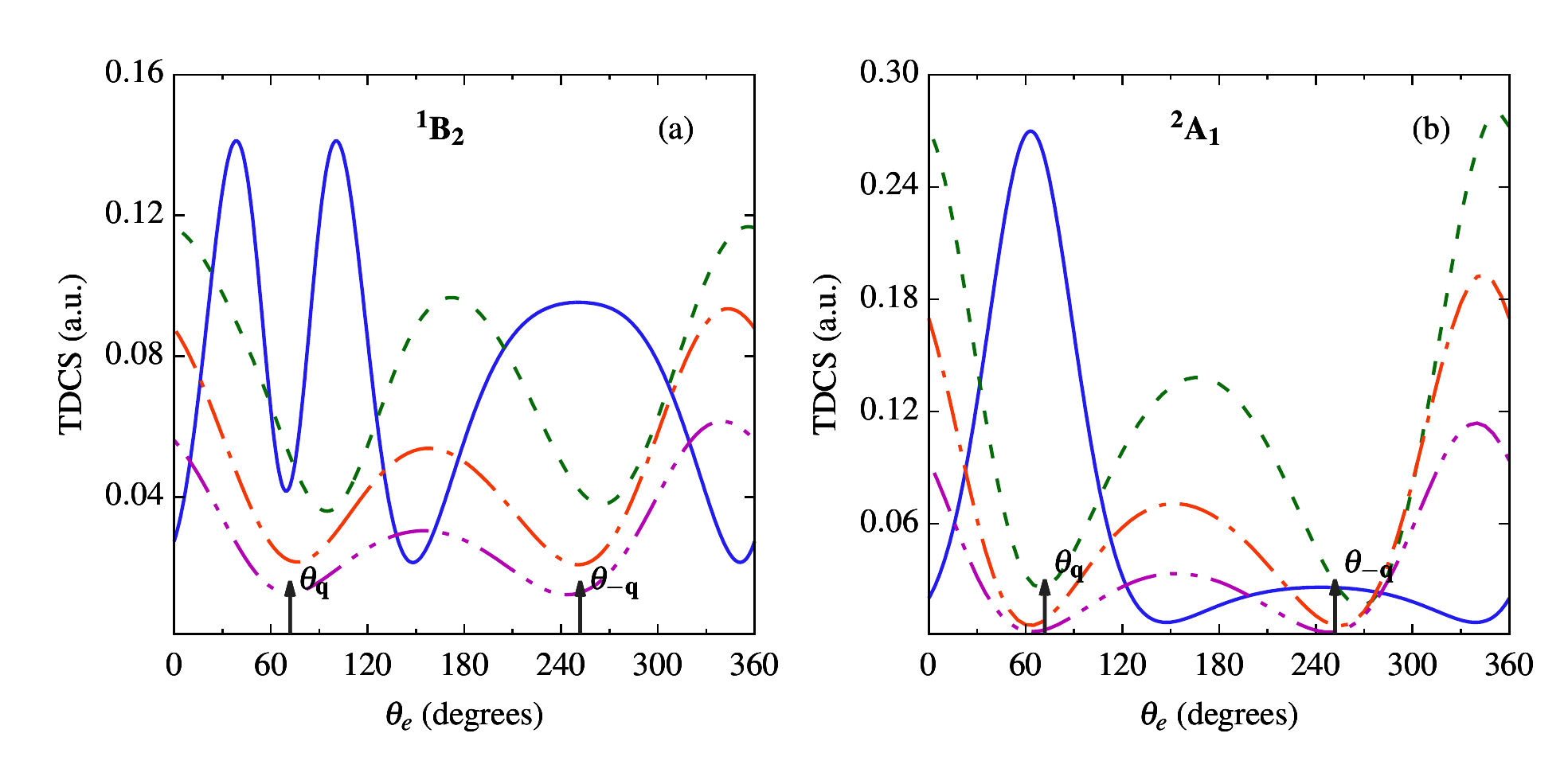}
\caption{Same as figure \ref{fig5},except here for both the $^1B_2$ and $^2A_1$ orbitals and $E_e$ = 10eV. The results for {\it m} $\neq$ 0 are scaled up by a factor of 2 in the sub-figure(a) and by 10 in the sub-figure(b).}\label{fig6}
\end{figure*}

\begin{figure}[htp]
\includegraphics[width=1.0\columnwidth]{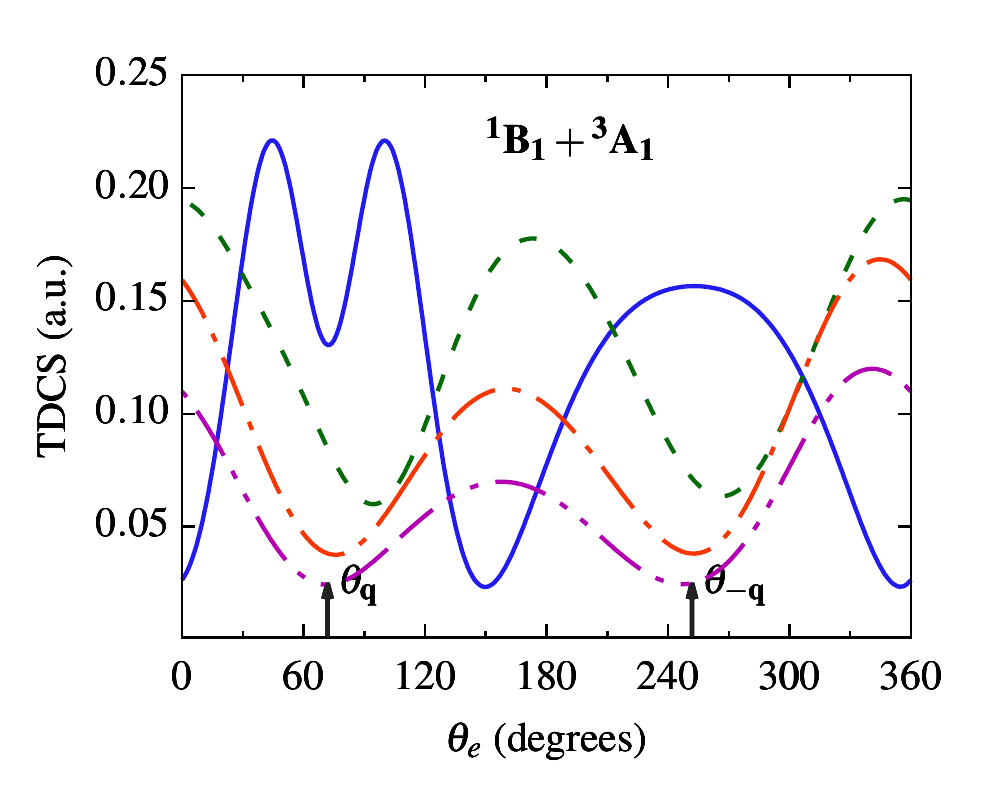}
\caption{Summed TDCS of the $^1B_1$ and $^3A_1$ orbitals as a function of ejection angle $\theta_e$ for a twisted electron beam. Kinematics is same as mentioned in figure \ref{fig3}. The magnitude of the TDCS for {\it m} $\neq$ 0 is scaled up by a factor of 2.}\label{fig7}
\end{figure} 
 
 	The molecular wave-function for the different orbitals of the water molecule is constructed from the Linear Combination of Atomic Orbitals (LCAO). Thus, the overall behavior of the TDCSs depend on the primary atomic component of each molecular orbital. As mentioned earlier, the $^1B_1$ orbital’s character is  essentially dictated by a $2p_{+1}$ atomic orbital, the $^3A_1$ by $2p_0$, the $^1B_2$ by $2p_{-1}$ and the $^2A_1$ by $2s$ \cite{Hafid1993, Hanssen1994, Champion2001}. For the present kinematics, the ionization process reveals these features in the angular distribution of the TDCS vs. $\theta_e$. The two-peak structure around the binary region and the single peak in the recoil region for the orbitals $^1B_1$, $^1B_2$ and $^3A_1$ is due to the strong {\it p}-like character of the orbitals (see figure \ref{fig3} (a)-(c)). While, due to the atomic {\it s}-like character of the $^2A_1$ orbital, the binary peak is present along the momentum transfer direction and recoil peak along the recoil direction \cite{Champion2001, Sanctis2015}. From figure \ref{fig3}, we observe that our theoretical model reproduces the experimental results quite well. 
 	
		In figure \ref{fig4}, we present the angular profile for the summed contributions from the individual $^1B_1$ and $^3A_1$ orbitals since, due to low energy resolution, the experiments are not able to resolve the peaks, particularly in the recoil region \cite{Milne2004}. The two-peak structure demonstrates the atomic {\it p}-like character associated with both the orbitals. Our theoretical calculation reproduces the angular profile in the binary region quite well.  Since the experimental results are on a relative scale, we have normalized them to compare the results with our theoretical results in the binary peak region.
		
\begin{center} 	
\begin{figure*}
\includegraphics[width=2.0\columnwidth]{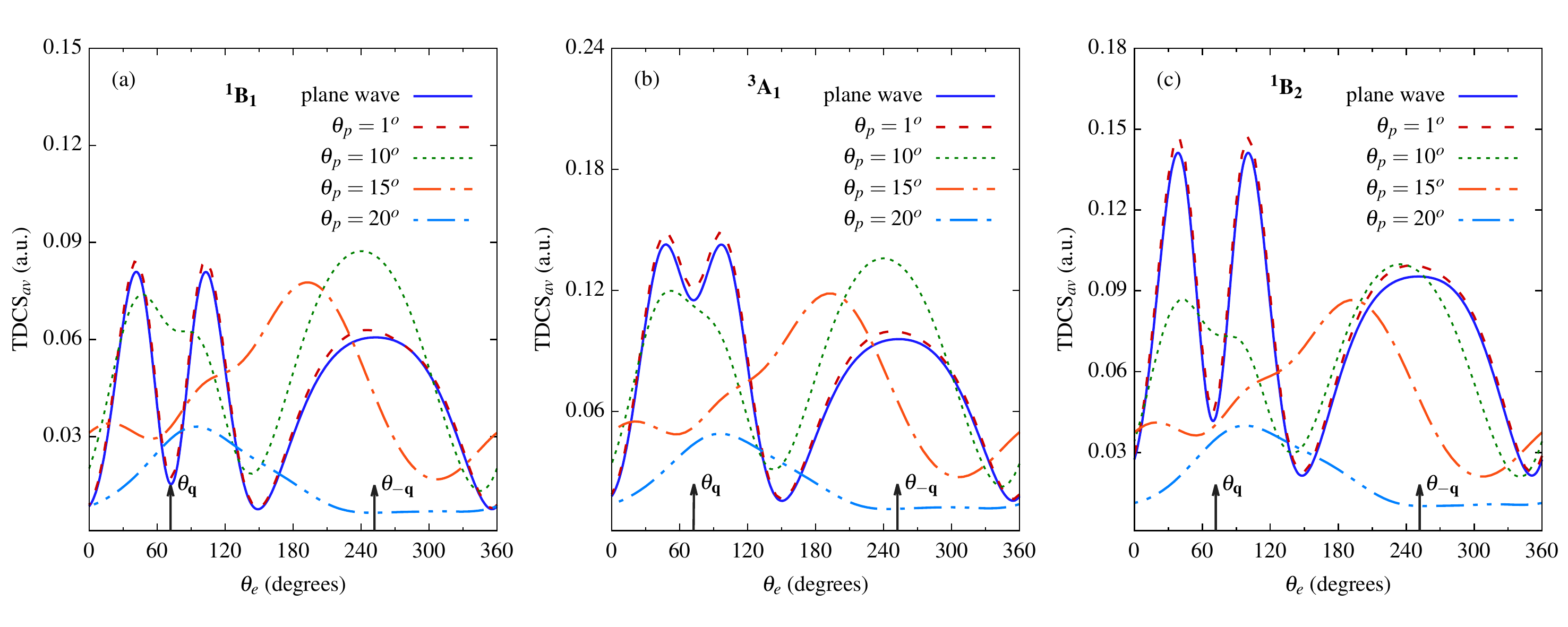}
\caption{(TDCS)$_{av}$ plotted as a function of the ejected electron angle $\theta_p$ for the plane wave (solid line) and  twisted electron beam for different opening angles as shown in the frames of each sub-figure. The kinematics is same as used in figure \ref{fig5} and \ref{fig6}. We have scaled down the (TDCS)$_{av}$ by a factor of 6 for the $^1B_1$ orbital for all the $\theta_p$s. For $^3A_1$ and $^1B_2$ orbitals, we have scaled down the magnitude by a factor of 1.5 for all cases.}\label{fig8}
\end{figure*}  
\end{center}
	
\subsection{Angular profiles of the TDCS for twisted electron wave}

In this section, we present the results of our calculation of the  TDCS with the twisted electron beam for $E_i$ = 250eV, $E_e$ = 10eV (8eV for $^3A1$ orbital) and $\theta_s$ = 15\textdegree \ in the co-planar asymmetric geometrical mode. Here, we keep $\theta_p$ = $\theta_s$ and vary the OAM number {\it m} from 1 to 3 in a step of 1. In figures \ref{fig5} - \ref{fig7}, the blue solid, green dashed, orange dashed-dotted and magenta dashed-dotted-dotted  curves represent the results for plane wave, {\it m}  = 1, 2 and 3 respectively. Figure \ref{fig5} (a) and (b) represent the TDCS for the $^1B_1$ and $^3A_1$ orbitals, while figure \ref{fig6} (a) and (b) represent that for the $^1B_2$ and $^2A_1$ orbitals respectively. As can be seen from the figures \ref{fig5} and \ref{fig6},  the magnitude of the TDCS for {\it m} $\neq$ 0 is reduced when compared with the {\it m} = 0 (the plane wave calculation). We have multiplied the results for all {\it m} for $^1B_1$, $^1B_2$, $^3A_1$ orbitals by a factor of 2 and by a factor of 10 for $^2A_1$ orbital. The magnitude further reduces when we gradually increase {\it m} upto 3 (see magenta dashed-dotted-dotted curve in figure \ref{fig5} and \ref{fig6}).

 	For the {\it p} dominant orbitals, {\it i.e.} $^1B_1$, $^1B_2$, $^3A_1$, we observe that the two-peak structure around the binary region disappears for all {\it m} (see dashed, dashed-dotted and dashed-dotted-dotted curves in the region marked by arrow in the binary region in figure \ref{fig5}(a), (b) and \ref{fig6}(a)). For the three orbitals, we observe a prominent contribution in the TDCS in the forward and backward direction for {\it m} = 1,2 and 3 (see peaks around $\theta_e$ = 0\textdegree (360\textdegree) and 180\textdegree for dashed, dashed-dotted and dashed-dotted-dotted curves).  For the $^2A_1$ orbital, for {\it m} = 1, 2 and 3 we observe substantial contribution in the forward and backward regions (see peaks around $\theta_e$ = 0\textdegree (360\textdegree) and 180\textdegree \ for dashed, dashed-dotted and dashed-dotted-dotted curves in figure \ref{fig6}(b)).
In both the figures \ref{fig5} and \ref{fig6}, we observe a minimum at/around the plane wave linear momentum transfer direction (see dashed, dashed-dotted, and dashed-dotted-dotted curves in figure \ref{fig5} and \ref{fig6} around the arrows). Due to an additional transverse momentum component in the incident momentum for the twisted electron beam, the peaks observed for the plane wave case are shifted significantly for the twisted electron case.

	In figure \ref{fig7}, we present the angular profiles of TDCS for the summed contributions of the $^1B_1$ and $^3A_1$ orbitals as a function of ejected electron angle $\theta_e$ for the twisted electron beam. The two-peak structure, as observed for the plane wave, disappears for the twisted electron beam as well. We also observe peaks in the forward and backward direction (see peaks around $\theta_e$ = 0\textdegree and 180\textdegree) as observed in the earlier cases. We have scaled up our calculations for the twisted electron beam by a factor of 2. We found that with an increasing {\it m}, the magnitude of the TDCS decreases.
In figures \ref{fig5}-\ref{fig7}, we observe that for an increasing OAM number {\it m}, the ratio of forward to backward peak increases. For example, for the $^1B_1$ orbital the ratio of the forward peak to backward peak for {\it m} =1, 2 and 3 is 1.126, 1.568 and 1.775 respectively (see dashed, dashed-dotted and dashed-dotted-dotted curves in figure \ref{fig5}(a)). Also, with an increasing {\it m}, both the forward and backward peaks shift towards the smaller angle (see dashed, dashed-dotted and dashed-dotted-dotted curves around $\theta_e$ = 180\textdegree in figure \ref{fig5}-\ref{fig7}). 

\subsection{Angular profiles for the (TDCS)$_{av}$ for a macroscopic $H_2O$ molecular target}
\begin{figure}[ht]
\includegraphics[width=1.0\columnwidth]{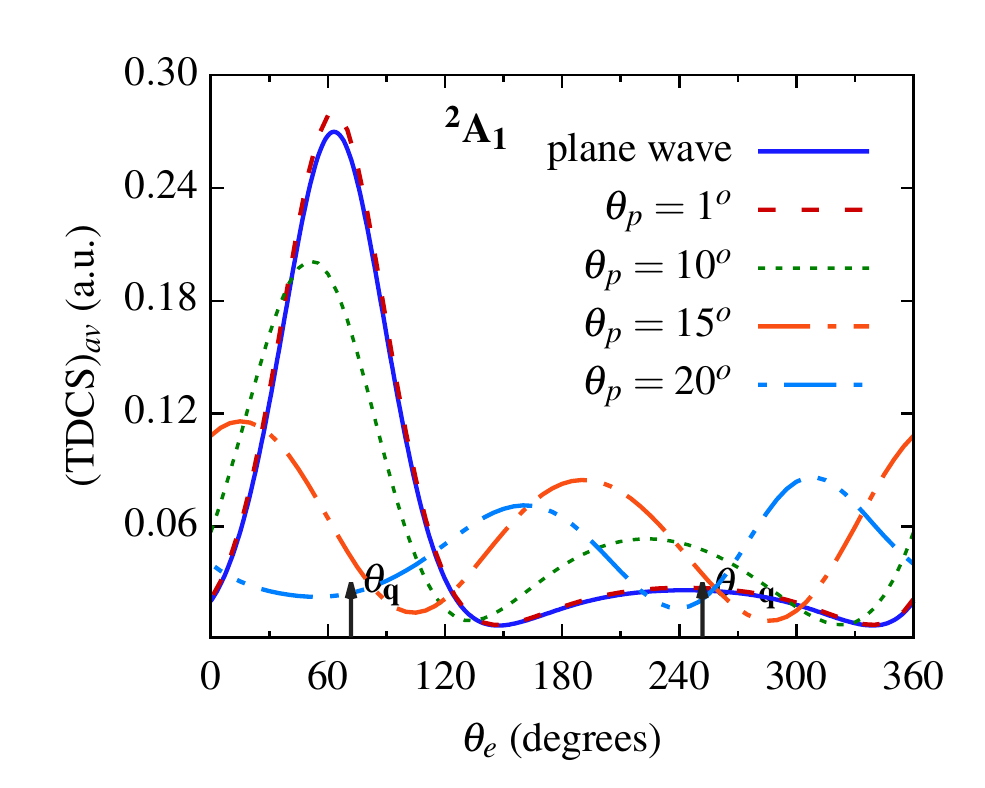}
\caption{Same as figure \ref{fig8}, except for the $^2A_1$ orbital and $E_e$ = 10eV. For $\theta_p$ = 1\textdegree \ the magnitude of the (TDCS)$_{av}$ is scaled down by a factor of 1.5, for $\theta_p$ = 15\textdegree \ and 20\textdegree \ the magnitude is scaled up by a factor of 2 and 5 respectively. We have not used any scaling factor for the $^2A_1$ orbital for $\theta_p$ = 10\textdegree.}\label{fig9}
\end{figure} 

In figure \ref{fig8} and \ref{fig9}, we present the results of our calculations for the TDCS averaged over the impact parameter \textbf{b}, (TDCS)$_{av}$, as a function of the ejected electron angle ($\theta_e$). The (TDCS)$_{av}$ depends only on the opening angle $\theta_p$ of the incident twisted electron beam (see equation \ref{21}). Figure \ref{fig8} and \ref{fig9} represent the (TDCS)$_{av}$ for the kinematics $E_i$ = 250eV, $E_e$ = 10eV (8eV for the $^3A_1$ orbital) and $\theta_s$ =  15\textdegree. We compare the results of the averaged TDCS ((TDCS)$_{av}$) with the plane wave results. We present the angular profiles for $\theta_p$ = 1\textdegree, 10\textdegree, 15\textdegree \ and 20\textdegree \ to compare them with that for the plane wave results. For a better comparison among these calculations, we have scaled down the (TDCS)$_{av}$ magnitude by a factor of 6 for the $^1B_1$ orbital for all the cases of $\theta_p$. For $^3A_1$ and $^1B_2$ orbitals, we have scaled down the magnitude by a factor of 1.5 for all cases. For the $^2A_1$ orbital, we have used different factors like,  for $\theta_p$ = 1\textdegree \ we scale it down by a factor of 1.5, for $\theta_p$ = 15\textdegree \ and 20\textdegree \ we scale it up by a factor of 2 and 5 respectively. We have not used any scaling factor for the $^2A_1$ orbital for $\theta_p$ = 10\textdegree. From figures \ref{fig8} and \ref{fig9}, we observe that for a smaller opening angle, like $\theta_p$ = 1\textdegree, the angular profile of the (TDCS)$_{av}$ is similar to that of the plane wave for all the orbitals (see solid and dashed curves in figure \ref{fig8} and \ref{fig9}). For $\theta_p$ = 10\textdegree, the characteristic two peak structure in the binary region for $^1B_1$, $^3A_1$ and $^1B_2$ orbitals disappear and we observe a broad single peak structure (see dotted curves in figure \ref{fig8}(a), \ref{fig8}(b) and \ref{fig8}(c) ). For the $^2A_1$ orbital, however, the binary peak structure is maintained (see dotted curve in figure \ref{fig9}). The recoil peak is, however, observed for all the orbitals for $\theta_p$ = 10\textdegree \ (see dotted curve around $\theta_{-\mathbf{q}}$ in figure \ref{fig8} and \ref{fig9}). We also observe that both the binary and recoil peaks are shifted from the plane wave momentum transfer direction (see dotted curves in figure \ref{fig8} and \ref{fig9}). For the $\theta_p$ = $\theta_s$ (15\textdegree) case, we observe a prominent peak in the backward region and a shallow peak in the forward region  for $^1B_1$, $^1B_2$ and $^3A_1$ orbitals while for the $^2A_1$ orbital we observe prominent peaks for both the forward and backward regions (see dashed-dotted curves in figure \ref{fig8}(a)-(c) and \ref{fig9} around $\theta_e$ = 0\textdegree \ and 180\textdegree). For a large opening angle, like $\theta_p$ = 20\textdegree, we observe a single peak in the angular profile of the (TDCS)$_{av}$  for the {\it p}-dominant orbitals ($^1B_1$, $^1B_2$ and $^3A_1$) while a two peak structure for the $^2A_1$ orbital (see dashed-dotted-dotted curves in figure \ref{fig8} and \ref{fig9}). In all the cases, we observe that the magnitude of the (TDCS)$_{av}$ decreases with an increasing opening angle ($\theta_p$).


\section{Conclusion}\label{sec4}
In this paper, we have presented the theoretical study of the triple differential cross-sections (TDCS) for an (e,2e) process on $H_2O$ molecule by the twisted electron beam. We studied the angular distributions of the TDCS for the co-planar asymmetric geometry in the first Born approximation for both the plane and twisted electron beam. We have studied the effect of the OAM number, {\it m}, on the TDCS for {\it m} = 1, 2 and 3. We have benchmarked our theoretical results with the experimental data for the plane wave electron beam. 
  For the (e,2e) ionization of the $H_2O$ by twisted electron impact, we observe that for the $^1B_1$,$^3A_1$ and $^1B_2$ orbitals (exhibiting atomic ``p''-type orbital characteristics), the two peak structures in the binary region, which is the signature of the {\it p}-type atomic orbital, disappear. For the $^2A_1$ orbital (governed by an atomic ``s''-type orbital) the peaks in the binary and recoil region are no longer present. We observe peaks in the forward and backward direction for all the outer orbitals in contrast to their plane wave ionization cross-section profiles. 
 We also observed that with an increasing value of the OAM number {\it m}, the magnitude of the TDCS decreases. We also discuss the (TDCS)$_{av}$ (averaged over the impact parameter \textbf{b}) as a function of the opening angle $\theta_p$ of the twisted electron beam. For a macroscopic target, the angular profiles of (TDCS)$_{av}$ significantly depend on the opening angle ($\theta_p$) of the twisted electron beam.

	Our present communication is the first attempt to investigate the (e,2e) process on the $H_2O$ molecule to unravel the effects of the twisted electron’s different parameters on the angular profile of the TDCS. We have used the 1CW wave function in our theoretical model to study the TDCS. The present study can also be extended for other molecular targets, like $N_2$, $NH_3$, $CH_4$ etc.  Besides this,  one can further explore the differential cross-sections going beyond the first Born approximation. To our knowledge, no comprehensive research on the (e,2e) processes on the $H_2O$ molecule for the twisted electrons has been done. Hence, our findings in this work must be taken from that perspective. We are confident that the present work will help to progress theoretical and experimental research in this subject. The present study can be further extended for a twisted electron beam impact ionization using more sophisticated models, like DWBA, 2CW, BBK, and DS3C, in near future. \cite{Champion2006, Ren2017, Gong2018, Psingh2019}.
 
\section*{Acknowledgments} Authors acknowledge Didier S\'{e}billeau for his help in our code development for the computation of TDCS.

%

\end{document}